\documentclass{iau} 
\usepackage{graphicx}

\title[Dwarf Galaxies in the Core of Coma Cluster] 
{Dwarf Galaxies in the Core of Coma Cluster}

\author[S N Hasan, Priya Shah and P Nagamani]   
{S N Hasan$^1$, Priya Shah$^2$ and P Nagamani$^3$}

\affiliation{$^1$Department of Mathematics, Maulana Azad National Urdu University, 
Gachibowli,\\ Hyderabad 500 032, India \\ email: {\tt hasan.najam@gmail.com} \\ [\affilskip] $^2$Department of Physics, Maulana Azad National Urdu University, 
Gachibowli, \\Hyderabad 500 032, India \\ email: {\tt priya.hasan@gmail.com} \\[\affilskip]
$^3$Department of Astronomy, Osmania University, Hyderabad 500 007, India}

\pubyear{2018}
\volume{344}  
\jname{Dwarf Galaxies: From the Deep Universe to the Present}
\editors{Kristen McQuinn and Sabrina Stierwalt, eds.}
\begin{document}

\maketitle

\begin{abstract}

 Dwarf galaxies constitute 18\% of the galaxies in the core of the Coma Cluster. We present the correlation between structural properties and morphology of galaxies in the central region of Coma Cluster for 221 objects within the apparent magnitude range m $<$ 19.5. The data is taken from the HST/ACS Coma Cluster Treasury Survey.  For cluster membership we have used photometric redshifts  and spectroscopic redshifts from literature. From the investigation of correlations of effective radius, Sersic index, absolute magnitude and bulge to total light ratio, we find the galaxies are distributed as follows: dwarfs 18\%,  E/SO 33\%, SO 22\%, Sb \& Sb0 17\% and  10\% are Spirals+Irregulars+Ring. We found that multiple component fits are best for giants and the single Sersic fit is best for dwarfs \& spiral galaxies. We shall try to explain why the single Sersic fit is best for dwarfs and what kind of stellar orbits do they correspond to using the bulge Sersic index of dwarfs . 
\end{abstract}

\firstsection 

\section{Introduction}
Morphological studies of galaxies are a useful starting point to understand the structure and formation of galaxies. \cite[Buta(2013)]{buta13} has an excellent review that has an opening statement that `galaxy morphology has many structures that are suggestive of various processes or stages of secular evolution'. 

We present the morphology and structural properties  of galaxies in the central region of Coma Cluster within 0.5 Mpc (0.3 Degree). For cluster membership, we have used the photometric and spectroscopic redshifts  from \cite[Mobasher et al.(2001)]{2001ApJS..137..279M}, SDSS  DR9 \cite[Ahn et al.(2012)]{2012ApJS..203...21A}  and  \cite[Mahajan et al.(2011)]{2011MNRAS.412.1098M}. The main objective is that to study the morphology and correlation between structural properties of dwarf galaxies.


The Coma cluster was the target of an HST-ACS Treasury program designed for deep imaging in the passbands F475W and F814W. We have taken publicly available data from HST/ACS Coma Cluster Treasury survey data release 2.1, during 2006 November and 2007 January \cite[Hammer et al. (2010)]{hammer10}. This survey is a deep two passband  imaging survey of one of the nearest rich clusters of galaxies. 
 
The survey was designed to cover an area of 740 arcmin$^2$ in regions of varying 
density of both galaxies and intergalactic medium within the cluster. Due to the ACS failure of January 27th 2007, it is only  28\% complete. Predicted survey depth for 10σ detections for optimal photometry of point sources is 27.6 in the F475W filter, and 26.8 mag in F814 (AB magnitudes). 

We did multiple component decomposition using GALFIT \cite[Peng et al. (2002)]{Peng_etal02} and \cite[Peng et al. (2010)]{Peng_etal10}. We also did a visual inspection of residuals of galaxies, and tabulated apparent magnitude, absolute magnitude, effective radius, b/a, and position angle are  given the effective morphology (Hasan, 2007).
GALFIT is a data analysis algorithm that fits 2-D analytic functions to galaxies and point sources directly to digital images.The functions used are  exponential, Sérsic/deVaucouleurs, Nuker, Gaussian, King, Moffat, and PSF. The PSF function is provided by the user. It also allows for simultaneous fitting of arbitrary number of components and combination of the above functional forms.

\begin{figure}[b]
\begin{center}
\includegraphics[width=3.4in]{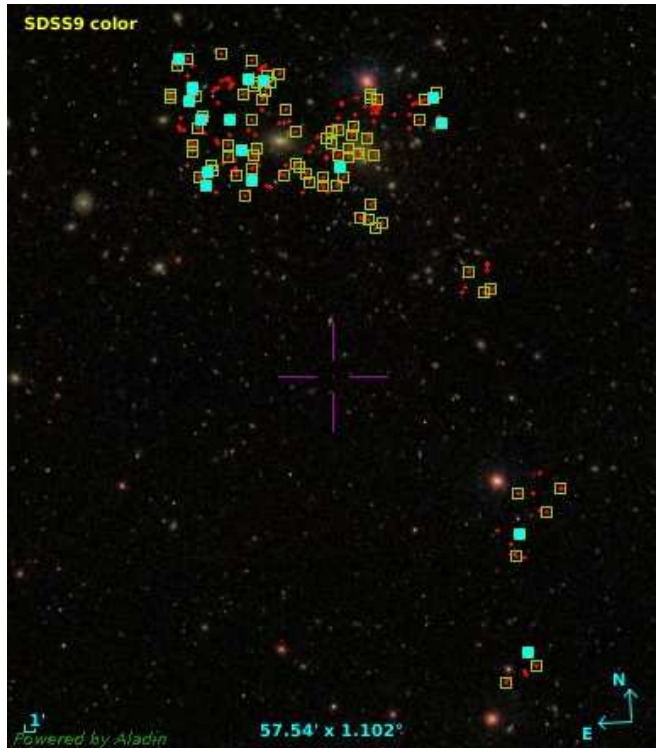} 
\caption{The HST/ACS Survey region of Coma Cluster. Double Sersic fits done for all 221 galaxies marked in red pluses, green squares are members and cyan squares are dwarfs.}
\label{f1}
\end{center}
\end{figure}

Double Sersic fits done for all 221 galaxies marked in red pluses, green squares are members and cyan squares are dwarfs (Fig. \ref{f1}).

We separate the disc and bulge components to study the nature of the bulge. When the Sersic index of the bulge $n < 2$, they are disk-like, psuedo bulges like in  spiral galaxies. They trace stars that are not orbiting randomly, but rather orbit in an ordered fashion in the same plane as the stars in the outer disk, implying secular evolution. If the Sersic index $n > 2$,they are classical bulges, found in elliptical galaxies, of older Population II stars,  in  random orbits. This would be the result of collisions of smaller structures. Agitated gravitational forces and torques disrupt the orbital paths of stars, resulting in the randomised bulge orbits.

We also categorized the galaxies based on their bulge to total ratios(given in brackets), where  E/S0 (0.8-1), S0(0.5-0.8), Sab(0.15-0.5), Sbc(0-0.15), Sd(0).

\section{Conclusions}
We found the following distribution of galaxies: dwarfs 18\%,  E/SO 33\%, SO 22\%, Sb \& Sb0 17\% and  10\% are Spirals+Irr+Ring. We found that multiple component fits are best for giants and the single Sersic fit is best for Dwarfs \& Spiral galaxies.
Dwarfs are not a new or different class of galaxies. Rather, they are the extreme products of transformation processes that get more important as gravitational potential wells get more shallow. (Kormendy, 2014). 

Multiple component fits are best for giants and the single Sersic fit is best for Dwarfs \& Spiral galaxies. Is that linked to the formation process? Dwarfs have $n < 2$, psuedo bulges, which imply ordered stellar orbits, with secular evolution. While is it because giants are built partially due to secular evolution and partially due to merging, that they have classes bulges  $n > 2$, and a linked formation process?

{}

\end{document}